\documentclass[dvips]{article}
\usepackage{graphics}
\usepackage{epsfig}
\usepackage{latexsym}
\usepackage{amsfonts}
\usepackage{amsmath}
 \usepackage{acronym}
\usepackage{graphics}
\newcommand{\be}{\begin{equation}}
\newcommand{\ee}{\end{equation}}
\newcommand{\bea}{\begin{eqnarray}}
\newcommand{\eea}{\end{eqnarray}}

\makeatletter
\@addtoreset{equation}{section}

 \makeatother
\begin{document}
\normalsize
\title{Whether the vacuum manifold in the Minkowskian non-Abelian model quantized by Dirac can be described with the aid of the superselection rules? }
\author
{\bf L.~D.~Lantsman.\\
  18109, Rostock, Germany; Mecklenburger Allee, 7\\ llantsman@freenet.de\\ Tel. (049)-0381-7990724.}
 \maketitle
 \begin{abstract}
  We intend to show that the vacuum manifold inherent in the Minkowskian non-Abelian model involving Higgs and Yang-Mills BPS vacuum modes and herewith quantized by Dirac can be described with the help of the superselection rules if and only if the ``discrete'' geometry for this vacuum manifold is assumed (it is just a necessary thing in order justify the Dirac fundamental quantization scheme applied to the mentioned model) and only in the infinitely narrow spatial region of the cylindrical shape where topologically nontrivial vortices are located inside this discrete vacuum manifold. 
 \end{abstract}
 \newpage 
In the recent paper \cite{disc} it was argued that the so-called Dirac {\it fundamental} quantization \cite{Dir} of the Minkowskian non-Abelian model involving Higgs and Yang-Mills (YM)  vacuum BPS modes, coming to the Gauss-shell reduction of the mentioned model in terms of topological Dirac variables, gauge invariant and transverse functionals of YM fields  \footnote{As important ``milestones'' in development of this model, it is worth to mention the papers \cite{Pervush1,David2,David3, Pervush2,LP1,LP2}. For the history of the question see also the survey \cite{fund}.}, is compatible with assuming the  ``discrete'' geometry for the appropriate vacuum manifold: 
\be
 \label{RYM}
 R_{\rm YM}= {\bf Z}\otimes G_0/U_0. \ee
 Such representation for the vacuum manifold $R_{\rm YM}$ is the direct consection of the ``discrete'' representations 
 \be
 \label{fact2}
SU(2)\simeq G_0\otimes {\bf Z}; \quad     U(1) \simeq U_0 \otimes {\bf Z}
 \ee
 for the initial, $SU(2)$, and residual, $U(1)$, gauge symmetries groups (respectively) in the Minkowskian non-Abelian Higgs model (we shall refer to this model as to the YMH model henceforth in the present study). 
 
 From the topological viewpoint, the  discrete representation (\ref{fact2}) for the gauge groups $G$ and $H$ extracts "small" (topologically trivial) and "large" (corresponding to topological numbers $n\neq 0$) gauge transformations in the complete set of appropriate gauge transformations (the idea of such subdividing for gauge transformations was suggested in Ref. \cite {Fadd2}).\par
 According to the terminology \cite {Fadd2}, the complete groups $G_0$ and $H_0$ just contain "small" gauge transformations,  that implies 
 \be
 \label{small}
\pi_n G_0 = \pi_n H_0 =0 \ee
for loops in the group spaces $G_0$ and $H_0$ in all the dimensions $n\geq 1$.\par 
Simultaneously, in definition,
\be
 \label{oneconnect} 
 \pi_0 G_0 = \pi_0 H_0 =0,
\ee
i.e. $G_0$ and $H_0$ are \it maximal connected components \rm (in the terminology \cite {Al.S.}) in their gauge groups (respectively, $G$ and $H$).\par
Later Eq. implies \cite {Al.S.} that
\be\label{dl} \pi_0 [G_0 \otimes {\bf Z}]= \pi_0 [G_0 \otimes {\bf Z}]=\pi_0 ({\bf Z})= {\bf Z}.\ee 
\medskip

It becomes obvious from Eq. (\ref {RYM}) that the  "small" coset $ G_0/U_0$ is one-connected:
$$\pi_1 (G_0/U_0)=0.$$
  Really, the coset $ G_0/U_0$ is treated as the space of $U_0$-orbits on $G_0$;  the latter space is one-connected. \par
One can see also the  topological equivalence between $ G_0/U_0$ and the subset of  one-dimensional ways on $R_{\rm YM}$ which can be contracted into a point.\par
\medskip
The vacuum manifold $ R_{\rm YM}$ is transparently multi-connected (i.e. {\it discrete}):
  \be
 \label{domena}
\pi_0 (R_{YM})={\bf Z}.
\ee
This implies \cite{Al.S.} that {\it  domain walls} exist between different   topological sectors in the Minkowskian Higgs model with vacuum BPS monopoles quantized by Dirac. \par
 The origin of said domain  walls is in the "discrete" factorisation (\ref{fact2}) of the residual gauge symmetry group $U(1)$.
 
 \medskip
As it is well known (see e.g. \S 7.2 in \cite{Linde} or the paper \cite{Ph.tr}), the width of a domain (or \it Bloch\rm, in the terminology  \cite{Ph.tr})  wall is roughly proportional to the inverse of the lowest mass among all the physical particles presented in the (gauge) model considered. \par
In Minkowskian Higgs models (without quarks) the typical such scale is the (effective) Higgs mass $m/\sqrt\lambda$. 
In particular, in the Minkowskian YMH model \cite{Pervush1,David2,David3, Pervush2,LP1,LP2} with vacuum BPS monopoles quantized by Dirac, $m/\sqrt\lambda$ is the only mass scale different from zero (in the ``world with quarks'' this remains almost correctly at assuming \cite{disc} $ m_0\ll m/\sqrt\lambda $ for any ``bare'' flavour mass $m_0$). 

 \medskip Together with the ``effective Higgs mass'' $m/\sqrt\lambda$, it is possible to write down the value roughly its inverse, i.e. having the length dimension. It is the  (typical) size $\epsilon$ of BPS monopoles.

It can be given as  \cite{ David2,LP1,LP2}
\be 
\label{masa}  \frac{1}{\epsilon}=\frac{gm}{\sqrt{\lambda}}\sim \frac{g^2<B^2>V}{4\pi}, \ee
with $g$ being the YM coupling constant. 
 Thus $\epsilon$ is inversely proportional to the {\it infinite} spatial volume $V =\int d^3x$ occupied by the appropriate YMH field configuration.
 
 Indeed, as it was argued recently in Ref. \cite{rem3}, in the assymptotical freedom limit $g\to 0$, $\epsilon$ {\it can take any finite values} (due to the $0\times \infty$ uncertainty in that case). This means that walls between topological domains inside $R_{\rm YM}$ can be of finite wides $O(\epsilon (0))\neq 0$, at the origin of coordinates.

 The said allows to assert that $\epsilon$ disappears in the infinite spatial volume limit $V\to\infty$ and when the coupling constant $g$ is fixed, i.e. actually in the (infrared) confinement region, while it is maximal at the origin of coordinates (herewith it can be set $ \epsilon (0)$). 
This means, due to the above reasoning \cite{Ph.tr}, that walls between topological domains inside $R_{\rm YM}$ become {\it of a fixed typical wide}, $O(\epsilon (0))\neq 0$, at the origin of coordinates.

The fact $\epsilon(\infty)\to 0$ is  also meaningful.  This implies  actual merging  topological domains inside the vacuum manifold $R_{\rm YM}$, (\ref{RYM}), at the spatial infinity.
This promotes the infrared topological confinement (destructive interference) of Gribov "large" multipliers $ v^{(n)}({\bf x})$ in gluonic and quark Green functions in all the orders of the perturbation theory. 
The latter fact was demonstrated utilizing the strict mathematical language in Ref. \cite{Azimov} (partially these arguments \cite{Azimov}  were reproduced in Ref. \cite{fund}).

\bigskip
The nontrivial isomorphism  \cite{Al.S.}
\be \label{iso} \pi_1 ({R}_{\rm YM})= \pi_0 (H)\neq 0\ee
correct \cite{disc} for the vacuum manifold $R_{YM}$, (\ref{RYM}) \footnote{It is the particular case of  the general relation \cite{Al.S.} 
$$  \pi_i (K)=  \pi_i (L_1)+\dots + \pi_i (L_r) 
    $$ 
for a group $K$ which is the  product of the groups $L_1 \dots L_r$ at a fixed $i$ (it is correctly for  the Lie groups of the series $SU$, $U$ and $SO$, with which modern theoretical physics deals).
}, implies the presence of {\it thread  topological defects} inside this manifold. \par 

As it was argued in the paper \cite{disc} (with the aid of the arguments \cite{Al.S.}),  this kind  of topological defects in the  Minkowskian YMH model \cite{Pervush1,David2,David3, Pervush2,LP1,LP2} with vacuum BPS monopoles quantized by Dirac can be represented by specific solutions in its YM and Higgs sectors: so-called (topologically nontrivial) threads. 

In particular, {\it in the Higgs sector} of the  Minkowskian YMH theory \cite{Pervush1,David2,David3, Pervush2,LP1,LP2} there are \cite{Al.S.} z-invariant (vacuum) Higgs solutions in a (small) neighbourhood of the origin of coordinates ($\rho \to 0$):
\be \label{Higgs-teta}
  \Phi^{(n)} (\rho, \theta, z)= \exp (M\theta)~ \phi  (\rho) \quad (n\in {\bf Z}), \quad \nabla_\mu \phi(\rho) \leq {\rm const}~\rho ^{-1-\delta}; \quad  \delta>0; \quad n\in {\bf Z};\ee
 $\rho=\sqrt{x^2+y^2} $ is the distance from the axis $z$.\par
 One claims for  Higgs thread solutions $\Phi^{(n)} (\rho, \theta, z)$ to join contineously and smoothly the vacuum Higgs BPS monopoles, belonging to the same topology $n$ and disappearing \cite{David3} at the origin of coordinates.  
 Herewith, speaking "in a smooth wise", we imply that the covariant derivative $D\Phi$ of any vacuum Higgs field $\Phi_a^{(n)}$ merges with the covariant derivative of such a vacuum Higgs BPS monopole solution. \par
 \medskip  The requirement for vacuum Higgs fields $\Phi_a^{(n)}$ to be smooth is quite natural if the goal is pursued, as it is done   in the  Minkowskian YMH model \cite{Pervush1,David2,David3, Pervush2,LP1,LP2} with vacuum BPS monopoles quantized by Dirac, to justify various rotary effects inherent in this model. \par

In particular, vacuum "electric" monopoles \cite{David2} \footnote{They involve, firstly, the topological varible $N(t)$ (with its time derivative $\dot N(t)$) introduced \cite{Pervush2} via the vacuum Chern-Simons functional 
\bea
\nu[A_0,\Phi^{(0)}]&=&\frac{g^2}{16\pi^2}\int\limits_{t_{\rm in} }^{t_{\rm out} }
dt  \int d^3x F^a_{\mu\nu} \widetilde{F}^{a\mu \nu}=\frac{\alpha_s}{2\pi}  \int d^3x F^a_{i0}B_i^a(\Phi^{(0)})[N(t_{\rm out}) -N(t_{\rm in})]\nonumber \\
 &&  =N(t_{\rm out}) -N(t_{\rm in})= \int\limits_{t_{\rm in} }^{t_{\rm out} } dt \dot N(t); \quad t_{\rm in}\to -\infty,~~ t_{\rm out}\to \infty;\nonumber
 \eea 
 and secondly, the real, i.e. {\it physical}, topological momentum 
 $$ P_N ={\dot N} I= 2\pi k +\theta; \quad \theta  \in [-\pi,\pi]. $$
}
\be 
\label{se} F^a_{i0}\equiv E_i^a=\dot N(t) ~(D_i (\Phi_k^{(0)})~ \Phi_{(0)})^a= P_N \frac {\alpha_s}{4\pi^2\epsilon} B_i^a (\Phi _{(0)})= (2\pi k +\theta) \frac {\alpha_s}{4\pi^2\epsilon} B_i^a(\Phi_{(0)}); \quad k\in {\bf Z};
       \ee 
       $$\alpha _s=\frac{g^2}{4\pi (\hbar c)^2 }; $$
       prove to be directly proportional to $D_i (\Phi_k^{(0)})~ \Phi_{(0)}$. 
       
       These vacuum "electric" monopoles, in  turn, enter explicitly the action functional 
       \be \label{rot} W_N=\int d^4x \frac {1}{2}(F_{0i}^c)^2 =\int dt\frac {{\dot N}^2 I}{2},\ee
  implicating the  ``rotary momentum'' \cite{David2}
\be \label{I} I=\int \sb {V} d^3x (D_i^{ac}(\Phi_a^{(0)})\Phi_{(0)c})^2 =
\frac {4\pi^2\epsilon(\infty)}{ \alpha _s}
=\frac {4\pi^2}{\alpha _s^2}\frac {1}{ V<B^2>}    \ee 
 and describing, in the Dirac fundamental quantization scheme \cite {Dir}, collective solid rotations inside the Minkowskian BPS monopole vacuum. 
 
  \medskip
Such (smooth) sawing together  appropriate vacuum Higgs modes $\Phi^{(n)}$ (which are \cite{Al.S.} specific thread rectilinear vortices) and BPS monopoles serves to remove the {\it seeming} contradiction between the manifest superfluid properties of the Minkowskian BPS monopole vacuum (suffered the Dirac fundamental quantization \cite {Dir}), setting by the Bogomolny'i \cite{LP1,LP2,Al.S.}, 
\be
\label{Bog}
 {\bf B} =\pm D \Phi, 
\ee
and {\it Gribov ambiguity} \cite{Pervush2,LP1,LP2},
\be
\label{Gribov.eq} [D^2 _i(\Phi _a^{(0)})]^{ab}\Phi_{(0)b} =0,
\ee
equations.

One can assert (following \cite {Pervush1}), and this can be seen from  (\ref{se}),  containing the vacuum ``magnetic'' field $\bf B$ given by the Bogomolny'i equation (\ref{Bog}), that, due to the Bianchi identity,
\be\label{colin} D~B\sim D~E =0  \ee
for vacuum "magnetic" and "electric" tensions: $\bf B$ and $\bf E$, respectively, in the quested YMH model \cite{Pervush1,David2,David3, Pervush2,LP1,LP2},  these tensions are, indeed, "transverse" vectors colinear each other.  
This just implies the potential nature of the "electric" tension $\bf E$, that can be perceived as the above contradiction, on the face of it. 

Going out from this contradiction seems to be just in locating (topologically nontrivial) threads in the infinitely narrow cylinder of the effective diameter $\epsilon(\infty)$ around the axis $z$ and in  joining (in a smooth wise) vacuum Higgs fields $\Phi_a^{(n)}$ and Higgs BPS monopole solutions (as it was explained in Ref. \cite{disc}).

In this case collective solid  rotations (vortices) inside the Minkowskian BPS monopole vacuum, occurring actually in that spatial region around the axis $z$ and described correctly by the action functional (\ref{rot}), become quite "legitimate", and simultaneously, the Gauss law constraint \cite{Pervush2}
\be \label{homo}   [D^2_i(\Phi ^{(0)})]^{ac} A_{0c}=0,        \ee 
just permitting, in the Minkowskian YMH model \cite{Pervush1,David2,David3, Pervush2,LP1,LP2} with vacuum BPS monopoles quantized by Dirac, the family of zero mode   solutions \cite{Pervush1, Pervush2} 
 \be \label{zero}  A_0^c(t,{\bf x})= {\dot N}(t) \Phi_{(0)}^c ({\bf x})\equiv Z^c,
   \ee
   generating ``electric monopoles'' (\ref{se}),  is satisfied outward  this  region with these smooth vacuum "electric" monopole solutions.  In  turn, one can refer \cite{disc} the ``electric monopoles'' (\ref{se})  to thread solutions since vacuum Higgs fields $\Phi_a^{(n)}$ are such.
   
 \bigskip   On the other hand, in the region of  thread topological defects inside the discrete vacuum manifold $R_{\rm YM}$, Eq.  (\ref{colin}) is violated since the vacuum "magnetic" field $\bf B$ suffers a break in this region. 
Really, according to the arguments \cite{BPS}, the  vacuum "magnetic" field $\bf B$ set via the Bogomol'nyi equation (\ref{Bog})  over YM and Higgs BPS monopole solutions diverges as $r^{-2}$ at the origin of coordinates. 

Simultaneously, following \cite{Al.S.},  thread ``counterparts'' of YM BPS monopole solutions $\Phi_i^{a{\rm BPS}}$ \cite{LP1,LP2}  can be constructed:
\be 
\label{Ateta}
  A  _\theta (\rho, \theta, z)=   \exp(iM\theta) A  _\theta (\rho) \exp(-iM\theta),    \ee 
with $M$ being the generator of the group $G_1$ of rigid rotations compensating changes in the vacuum YMH  ``thread'' configuration $(\Phi^a,A_\mu^a)$ (with $\Phi^a$ given in (\ref{Higgs-teta})) at rotations around the  axis $z$ of the chosen (rest) reference frame. 

In (\ref{Ateta}), 
\be \label{br}  A  _\theta (\rho) = M+ \beta (\rho),
  \ee
where the function $\beta (\rho)$ approaches zero as $\rho \to \infty$.  \par 

The elements of $G_1$ can be set as \cite{Al.S.} 
\be \label {gteta} g_\theta =\exp(iM\theta).    \ee
YM fields $A_\theta $ are manifestly invariant with respect to shifts along the axis $z$. 

Rectilinear threads $A_\theta $ don't coincide with vacuum YM BPS monopole solutions $\Phi_i^{a{\rm BPS}}$ \cite{LP1,LP2}, and, on the contrary, there are gaps between directions of "magnetic" tensions vectors: ${\bf B}_1$,
\be \label{B1} \vert {\bf B}_1 \vert \sim \partial _\rho A_\theta (\rho,\theta, z), \ee 
and $\bf B$, given by the Bogomol'nyi equation (\ref{Bog}) (and diverging as $r^{-2}$ at the origin of coordinates).

These gaps  testify in favour of the first-order phase transition \cite{disc} occurring in the Minkowskian YMH model \cite{Pervush1,David2,David3, Pervush2,LP1,LP2} with vacuum BPS monopoles quantized by Dirac. 

\medskip The important point of our above reasoning is that the vacuum expectation value of the Higgs field squared, $\sim <\Phi^a\Phi_a>$, cannot be treated as an order parameter in the Minkowskian YMH model \cite{Pervush1,David2,David3, Pervush2,LP1,LP2} with vacuum BPS monopoles quantized by Dirac. Otherwise, a flip should exist in the plot of a  Higgs field $\Phi^a(r)$ at the origin of coordinates, $r\to 0$, as a sign of the first-order phase transition occuring in the Minkowskian YMH model \cite{David2, David3, Pervush2,LP1,LP2}. But then it will be impossible to ``join'' continiously and smoothly  Higgs solutions $\Phi^a(r)$ with ``zero mode'' solutions
 $Z^a$ \cite{Pervush1}, (\ref{zero}), involving Higgs BPS monopole modes.  And this should contradict to the Dirac fundamental quantization of the model \cite{Pervush1,David2,David3, Pervush2,LP1,LP2}. 
 
Vice verse, the vacuum expectation value of the "magnetic" tension, $<B^2>$, can serve as an order parameter in the quested Minkowskian YMH model \cite{Pervush1,David2,David3, Pervush2,LP1,LP2}, with the  first-order phase transition taking place,  due to the obvious gap between directions of the "magnetic" tensions vectors ${\bf B}_1$ and ${\bf B}$ (such assumption was made already in Refs. \cite{LP1,LP2}, and then it was confitmed in the paper \cite{disc}).

This distinguish the  Minkowskian YMH model \cite{Pervush1,David2,David3, Pervush2,LP1,LP2} with vacuum BPS monopoles quantized by Dirac from another YM models (for instance, the 't Hooft-Polyakov model \cite{H-mon, Polyakov}) implying the  {\it continuous} $\sim S^2$ vacuum geometry, where just the vacuum expectation value of the Higgs field squared, \linebreak $<\Phi^a\Phi_a>$, serves as an order parameter). This is associated with the second-order phase transition taking place in such non-Abelian models (this was grounded, for example, in Ref. \cite{rem1} with the help of the arguments  \cite{Linde}).

\bigskip The first-order phase transition taking place in the Minkowskian YMH model \cite{Pervush1,David2,David3, Pervush2,LP1,LP2} with vacuum BPS monopole solutions quantized by Dirac comes \cite{disc} to the coexistence (in the absolute temperature limit $T\to 0$) of two thermodynamic phases inside the vacuum of that  model. These two thermodynamic phases are the phase of collective solid rotations, set by the action functional (\ref{rot}) (involving [topologically nontrivial] thread configurations $(\Phi^a,A_\mu^a)$ and generating ``electric monopoles'' $E_i^a$ \cite{David2}, (\ref{se})) and the phase of superfluid potential motions set by the Bogomol'nyi equation (\ref{Bog}) \cite{LP1,LP2,Al.S.} and the Gribov ambiguity equation (\ref{Gribov.eq}). 

The just described thermodynamic phases inside the Minkowskian YMH physical vacuum \cite{Pervush1,David2,David3, Pervush2,LP1,LP2} can be characterized by two different scales for the ``effective'' Higgs mass $m/\sqrt{\lambda}$.  For instance, collective solid rotations inside that vacuum correspond, as it is easy to see, to the zero mass scale $m/\sqrt{\lambda}\to 0$, while superfluid potential motions correspond to a nonzero mass scale $m/\sqrt{\lambda}\neq 0$. 

At $T\to 0$ the both thermodynamic phases inside the Minkowskian physical vacuum \cite{Pervush1,David2,David3, Pervush2,LP1,LP2} as if freeze  \cite{disc},  that gives a stable look to the studied model \cite{Pervush1,David2,David3, Pervush2,LP1,LP2}.  Nevertheless, it remains an important question,  in the framework of the first-order phase transition occurring therein, which of the enumerated thermodynamic phases ``belongs'' to the ``true'' and which to the ``false'' (metastable) vacuum?

\medskip In the present study we attempt to ground that, for all  that, collective solid rotations inside the Minkowskian physical vacuum \cite{Pervush1,David2,David3, Pervush2,LP1,LP2} relate to the ``true vacuum'', while superfluid potential motions therein relate to the ``false'' vacuum.

 The key point in  grounding the "superselection rules", we discuss in the present study, will be once again the ``discrete'' vacuum geometry (\ref{RYM}) \cite{disc} us assumed for the appropriate vacuum manifold $R_{\rm YM}$.

In the coordinate region 
\be \label{reg}
r=\sqrt{x^2+y^2}\to 0;\quad {\rm arbitrary}~~ z\ee 
of the Minkowski space (i.e. [infinitely] near the axis $z$ of the chosen rest reference frame),  the  vacuum manifold $R_{\rm YM}$,  (\ref{RYM}), consists of topological domains separated by  walls of the typical thickness $\epsilon(0)\neq 0$. 

In this case the assumption is quite permissible that topological sectors inside the vacuum manifold $R_{\rm YM}$ in the pointed spatial region can be identified with the {\it superselection sectors} [\it coherent Hilbert spaces\rm] (see e.g. \S 6.2 in \cite{Logunov}).  

 Indeed, to  accomplish such an identification, some  conditions would be observed. Note, first of all, that the term ``coherent spaces'' implies \cite{Logunov} constructing physical Hilbert spaces ${\cal H}_n$ ($n\in {\bf Z}$), which are, from the physical viewpoint,   quantum analogues of topological sectors inside $R_{\rm YM}$. In turn, in definition,  coherent Hilbert spaces ${\cal H}_n$ would consist of vectors describing pure quantum states and forming  irreducible representations of these ${\cal H}_n$. Only thereafter, the vacuum manifold $R_{\rm YM}$ can be represented (in the meanwhile, theoretically!) as \cite{Logunov}
 \be \label{cohr}
 \hbar R_{\rm YM}\simeq  \oplus {\sb n} {\cal H}_n,
 \ee
where all the ${\cal H}_n$ are mutually orthogonal (the Planck constant $\hbar$  indicates, may be formally, that this is, indeed, the quantum analogue of the vacuum manifold $ R_{\rm YM}$).

 Eq. (\ref{cohr}) reflects  also \cite{Logunov} identifying the gauge and topological charges. It is quite justified in the Minkowskian YMH model \cite{Pervush1,David2,David3, Pervush2,LP1,LP2} quantized by Dirac due to the nature of topological Dirac variables $\hat A^D$ \cite{David2,David3},
  \bea
\label{degeneration}
 \hat A_k^D = v^{(n)}({\bf x})T \exp \left\{\int  \limits_{t_0}^t d {\bar t}\hat A _0(\bar t,  {\bf x})\right\}\left({\hat A}_k^{(0)}+\partial_k\right ) \left[v^{(n)}({\bf x}) T \exp \left\{\int  \limits_{t_0}^t d {\bar t} \hat A _0(\bar t,{\bf x})\right\}\right]^{-1}; \quad D^k\hat A_k^D =0; 
\eea
$$k=1,2,3; $$ 
involving (``small'', ``large'') gauge matrices $v^{(n)} ({\bf x})$ \cite{Fadd2}.

\medskip The key point of the present reasoning is that each ${\cal H}_n$ consist of vectors describing pure quantum states. But as far as it is correctly for the vacuum manifold $R_{\rm YM}$? Obviously, in the light  identifying the gauge and topological charges in the Minkowskian YMH model \cite{Pervush1,David2,David3, Pervush2,LP1,LP2} quantized by Dirac,  each coherent physical Hilbert space ${\cal H}_n$ would imply fixing a definite topology $n$ inside $R_{\rm YM}$. Then one can speak about the pure quantum states sweeping ${\cal H}_n$.  These pure quantum states can be transformed  each into another by means of ``small'' gauge matrices $v^{(0)} ({\bf x})$; on the other hand,  there is a one-to-one correspondence between  this  ${\cal H}_n$ and the set of ``large'' gauge matrices $v^{(n)} ({\bf x})$. 

In the theoretical-group language, a one-to-one correspondence can be traced between a Hilbert space ${\cal H}_n$ and the appropriate ``small'' orbit of $U(1)\subset SU(2)$. The said allows, following Ref. \cite{WittenA51}, to represent a  (physical) coherent Hilbert space ${\cal H}_n$ as $V\otimes V_u$ ($u\in U(1)$), with $V$ being the  Hilbert space in the usual ``classical'' sence, while $V_u$ being the (finite-dimensional) vector space topologically equivalent to the n$^{\rm th}$ topological sector inside $U(1)\simeq S^1$ group space. 

\bigskip
There are, however, definite remarks and questions, whether and to which extend it is posible to do this fixing a definite topology inside the vacuum manifold $R_{\rm YM}$?

As it was discussed in Ref.  \cite{disc} repeating the arguments \cite{Al.S.}, YM fields with equal  magnetic charges ${\bf m}\neq 0$ can annihilate mutually at crossing  topologically nontrivial threads which are always present  inside the discrete  manifold $R_{\rm YM}$. Furthermore, topological deffects (hedgehogs and threads in the discussed YMH model \cite{Pervush1,David2,David3, Pervush2,LP1,LP2})  can merge and annihilate quite spontaneously, beyond the above colliding processes  (see e.g. \S$\Phi$1 in  \cite{Al.S.}). Also, due to finite domain walls between different topological sectors inside the vacuum manifold $R_{\rm YM}$ in the spatial region (\ref{reg}) near the origin of coordinates, an interaction between these sectors becomes quite possible.

All this, on the face of it, impedes fixing a definite topology  inside $R_{\rm YM}$ (as a result, quantum states become mixed).  But the  reasonable way out from this problem seems to be the following.  One consider all the  processes with merging and annihilating topological defects as those violating thermodynamic equilibrium inside  $R_{YM}$.  In this case it is possible to fix a  definite topology $n$ inside the discrete vacuum manifold $R_{\rm YM}$ and to construct the appropriate coherent physical Hilbert spaces ${\cal H}_n$ if the {\it relaxation time} $\tau$ during which merging and annihilating topological defects proceeds is large enough (see e.g. \S 110 in \cite{Landau5}). The same concerns also the interactions between different topological domains, i. e. domain walls. The latter can be interpreted in terms of "step voltage" between (neighboring) topological sectors, as it was discussed in the paper \cite{disc}. Then (quantum) fluctuations of physical parameters referring to $R_{\rm YM}$ will be small and these parameters will refer to a thermodynamic equilibrium.  Only at these assumptions one can assert that the vacuum manifold $R_{\rm YM}$ is in a pure  quantum state (corresponding to the direct sum $\oplus {\sb n} {\cal H}_n$).  As it was demonstrated in \cite{Landau5}, the above claim $\tau\to\infty$ is equivalent to the Gaussian distribution of physical parameters characterizing $R_{\rm YM}$.

On the other hand,  the knowledge about the free energy $F$ of the vacuum manifold $R_{\rm YM}$ is very important to decide whether  physical parameters characterizing $R_{\rm YM}$ are distributed Gaussian (that is equivalent to finding this manifold in a pure  quantum state). 

  The maximum entropy point of a model can be normalized to be \cite{Landau5} $S_{\rm max}=S\vert_{x=\bar{x}=0}$ (in our case $x$ is a physical parameter characterizing $R_{YM}$ while $\bar x$ is its [Gibbs] average). Whence 
  
  \be \label{24}
  \frac{\partial S}{\partial x} \vert_{x=0} =0; \quad    \frac{\partial^2 S}{\partial x^2} \vert_{x=0} <0.
  \ee
  Then in a neighborhood of $x=0$, the entropy $S=(E-F)/T$  inherent in the vacuum manifold $R_{\rm YM}$ can be expand in the series \cite{Landau5}
  \be \label{series}
  S(x)\sim S(0)-  \frac \beta 2 x^2; \quad \beta={\rm const}>0;
  \ee
  by the powers of $x$. 

In this case the probability $w(x)$ for $x$ to be in the interval $[x, x+dx]$ which is directly proportional to $e^{S(x)}$: 
\be \label{prob}
w(x)= {\rm const}\cdot e^{S(x)},
\ee
just results the Gaussian distribution for $x$:
\be \label{Gaussian}
w(x) dx=A e^{\frac {-\beta} 2 x^2} dx; \quad A=\sqrt{\beta/2\pi}.
\ee
We see thus the importance  knowing the complete Hamiltonian describing $R_{\rm YM}$, (\ref{RYM}).  In particular, it is worth to study the item in this  Hamiltonian responsible for colliding vacuum BPS monopole modes with (topologically nontrivial) threads (i.e. YM fields $A_\theta$ \cite{disc,Al.S.}, (\ref{Ateta})). It is optimal herewith the situation when $\beta$ is small. Then the entropy $S$ go to its maximum (that corresponds \cite{Landau5} to the minimum of the free energy $F$).

Thus for a system of (physical) fields it is energetically advantageous that corrections to the free energy $F$ conditioned by merging and annihilating topological defects are small and ``belong'' to the perturbation theory.

 In the framework of the Minkowskian YMH model \cite{Pervush1,David2,David3, Pervush2,LP1,LP2} quantized by Dirac, for    vacuum BPS monopole modes colliding \cite{disc,Al.S.} with (topologically nontrivial) threads, it is important, in the light of the said above, to understand whether it is described by a perturbation theory  in the YM effective coupling constant $\alpha_s$ (that corresponds to small values of the appropriate $\beta$) or  (although finding out the direct dependence $\beta$ on $\alpha_s$ is, apparently, a challenge). 
 
 If it is so, the arising radiative corrections result a shift of the ``true'' vacuum. This implies, in  turn, a ``blurring'' of the first-order phase transition picture taking place \cite{disc} in the Minkowskian YMH model \cite{Pervush1,David2,David3, Pervush2,LP1,LP2} quantized by Dirac.
 
\medskip  On the other hand,  setting $\bar x=(\bar x)^2=0$ refers rather to the symmetric ($SU(2)$) phase of the quested model. But our interest in the Minkowskian YMH model \cite{Pervush1,David2,David3, Pervush2,LP1,LP2} is its less  symmetrical ($U(1)$) phase, in which various vacuum superfluid and rotary effects are revealed (in the framework of the first-order phase transition picture). 

For example,  $x\neq 0$ (then $(\bar x)^2\neq 0$) can be  ordering parameter characterizing the Minkowskian YMH model \cite{Pervush1,David2,David3, Pervush2,LP1,LP2} quantized by Dirac (it is \cite{LP1,LP2} the  $\pm \sqrt{<B^2>}$ for the ``magnetic'' field squared ${\bf B}^2$). 

In this case $x$ has the nonzero dispersion 
\be \label{disp}
D x= <(x-\bar x)>^2= M(x^2)-(M x)^2 \neq 0
\ee
($M x$ is the mathematical, i.e. vacuum in the physical context, expectation value  of $x$).  Thinking that $M (x)=0$ (this is an ordinary  assumption in QFT), one has $D x= M(x^2)\equiv <x^2>$. 

On the other hand  \cite{Landau5}, now (at the assumption $M (x)\equiv <x>=0$) 
\be \label{29}
<(x-\bar x)>^2= <x^2>=\int \limits _{-\infty} ^ \infty x^2w(x)dx=\beta^{-1}.
\ee
Just this shows that the maximum of the entropy, corresponding to the limit $\beta\to 0$, can be achieved in the Minkowskian YMH model  \cite{Pervush1,David2,David3, Pervush2,LP1,LP2} quantized by Dirac if the minimum of the ordering parameter $ \sqrt{<B^2>}$ is absolute, i.e.  maximally possible deep. In other words, the maximal entropy (in the $T\to 0$ limit) is reached, obviously, over the ``true'' (absolute) vacuum, for which (in the majority of modern physical theories) $<B^2>\neq 0$. 

Such is the Minkowskian YMH model  \cite{Pervush1,David2,David3, Pervush2,LP1,LP2} quantized by Dirac, where, as in each Yang-Mills theory with the violated (initial) $SU(2)$ gauge symmetry,  $ \sqrt{<B^2>}$ serves as the gauge symmetry breaking parameter (and the order parameter simultaneously). The reason why namely  $ \sqrt{<B^2>}$ and not the VEV of the Higgs field squared serves as the order parameter in order parameter in the Minkowskian YMH model quantized by Dirac was explained in the papers \cite{disc,  LP2}.   As can be seen from Eq. (\ref{masa}), the Higgs (effective) mass $m/\sqrt{\lambda}$  (where $m$ and $\lambda$ are the Higgs mass and self-interacting constant, respectively) goes to infinity in the limit $V\to \infty$ at assuming that $<B^2>$ is finite in this limit. In this case \cite{LP1} the scalar (Higgs) field acquires an infinitely large mass and disappears from the spectrum
of physical excitations. Thus the role of the order parameter of the physical BPS monopole vacuum is "fixed" for $<B^2>$ in this infinite volume limit.

Indeed, how it becomes clear from the Bogomolny'i and Gribov ambiguity equations above, the nonzero VEV  $ \sqrt{<B^2>}$ is responsible for the superfluid properties of the YMH BPS monopole vacuum quantized by Dirac. And thus the ``true'' vacuum relates to the superfluid modes. 

In the paper \cite{disc} it was expected that in order the first order phase transition takes place in the Minkowskian YMH model  \cite{Pervush1,David2,David3, Pervush2,LP1,LP2} quantized by Dirac, it is necessary that 
\be \label{B1v}
<B_1^2> =0,
\ee
i.e. $$<\partial_ \rho A_\theta (\rho,\theta, z)> =0 \Leftrightarrow <\partial_\rho \beta(\rho)>=0 \Leftrightarrow  <\beta(\rho)>={\rm const}.$$ 
Thus the first order phase transition condition in model  \cite{Pervush1,David2,David3, Pervush2,LP1,LP2} can be reduced to the simple condition of (nonzero) VEV for $\beta(\rho)$ given in (\ref{br}).

At the obvious interpretation of the VEV of the magnetic field ${\bf B}_1$ squared as the order parameter for the rotary phase inside the Minkowskian YMH vacuum quantized by Dirac (besides that, the vectors ${\bf B}_1$ and $\bf B$ are not colinear, and this  testifies in favour of the  first order phase transition in the model us discussed), this allows us to assume \cite{disc} that the 
condition (\ref{B1v}) determines the {\it false vacuum} in the Minkowskian YMH model quantized by Dirac. Although only same system of differential equations involving the (quantized) fields entering us discussed model (in particular, the gauge potential $A_\mu^a$ responsible for rotary effects inside the Minkowskian YMH vacuum quantized by Dirac) can give the exact answer or the "rotary" vacuum identifiable with the condition (\ref{B1v}) is "false" indeed. For this aim, the knowledge about the explicit look of the item in the complete YMH Hamiltonian (Lagrangian) involving the "thread" configuration $(\Phi^a,A_\mu^a)$ \cite{Al.S.} is necessary. But it's beyond the present study.

 \medskip The case when  $x\neq 0$ ($<x>\neq 0$) is another parameter having a relation to the vacuum manifold $R_{\rm YM}$ is not a less  interesting. The one of such important parameters (along with $<B^2>$ discussed above) is  $m/\sqrt{\lambda}$  for the effective Higgs mass \cite{LP1,LP2}. It is obvious now that the $\beta\to 0$ limit (at which the  entropy $S$ of the vacuum manifold $R_{\rm YM}$ is maximum according to 
(\ref{series})) corresponds to the limit $m/\sqrt{\lambda}\neq 0$ (indeed, $m/\sqrt{\lambda}\to \infty$ as $V\to \infty $) for this parameter. 

\medskip Whence an interesting conclusion can be drawn that the maximum entropy in the Minkowskian YMH model \cite{Pervush1,David2,David3, Pervush2,LP1,LP2} quantized by Dirac is achieved  in the spatial region far from the origin of coordinates. On the other hand, this allows to apply the superselection rules \cite{Logunov} to this  manifold in order to construct the Hilbert space $\oplus {\sb n} {\cal H}_n$: in a definite sense, the latter one is a quantum analogue of $R_{\rm YM}$.  


\bigskip As it was  discussed in Ref. \cite{disc}, in the $r\to \infty$ limit, the ``geometrical'' picture of the vacuum manifold $R_{\rm YM}$ changes in a radical wise. Domain walls  become infinitely thin (as it can be seen from (\ref{masa})), and  this promotes merging topological domains inside $R_{\rm YM}$ in this spatial region. Note also that the large $r$ region (indeed, $r\sim ~{\rm 1~fm}$) is just the quarks, gluons confinement region for which the coupling constant $g\neq 0$, and this results $\epsilon(\infty) \to 0$ according to (\ref{masa}). In this case merging (annihilation)  topological defects cannot be considered as perturbation processes because of unsuppressed tunneling through such (infinitely) thin domain walls \footnote{The said resembles the visual picture when liquid helium II, possessing superfluidity, flows in parallel capillaries with   porous walls.
}. 

Namely  against this background of tunneling effects between topological domains inside $R_{\rm YM}$ at distances $r\gg 0$ superfluid potential motions proceed in the Minkowskian physical vacuum \cite{Pervush1,David2,David3, Pervush2,LP1,LP2} involving BPS monopole solutions and quantized by Dirac. \medskip

Thus to achieve the correct superselection description of the vacuum  manifold $R_{\rm YM}$, any coherent Hilbert space ${\cal H}_n$ would be {\it effectively} restricted in the Minkowskian {\it coordinate} space. The scalar product in such Hilbert spaces looks as following: 
\be \label{scalar product}
2\pi \int \limits_{0}^{r_1} f(r)g(r)r^2dr; \quad f(r),~ g(r)\in {\cal H}_n; \quad r=\sqrt{x^2+y^2}
\ee
(in cylindrical coordinates introduced in the Minkowskian space). The upper limit $r_1$ in the above Lebesgue integral can be evaluated as $r_1\sim O(\epsilon(0))$, i.e. it is enough small, but not zero. Note that the integration in the interval $[r_1,\infty]$ gives a vanishing contribution due to discussed above thinning and mutual mergering  topological domains at large distances.\medskip

It is easy to see (this, perhaps, will be done in the one of future studies the author plans) that ``thinning''  domain walls inside $R_{\rm YM}$  at distances $r\to\infty$ (with accompanying tunelling effects between topological domains inside this vacuum  manifold) promotes the infrared topological confinement in the spirit \cite{Azimov} i.e. surviving  only ``small'' Gribov multipliers  $v^{(n)} ({\bf x})$ in quark and gluonic Green functions in all the orders  of the perturbation theory. And it is the one of important gains of that ``thinning'', especially because such infrared topological confinement implies \cite{Pervush2} the confinement of gluons and quarks in the sense as it is realized ordinary in theoretical physic.

\bigskip As it was noted in Ref. \cite{disc} (and repeated again in the present study), the effective Higgs mass $m/\sqrt{\lambda}$ varies (in the Bogomolny'i limit  $m\to 0$, $\lambda\to 0$ \cite{LP1,LP2,Al.S.,BPS}) in the interval from  some finite value in the spatial region (\ref{reg}) locating (topologically nontrivial) thread configurations to the infinite value in the infrared limit $r\to\infty$. 
As it was shown in Refs. \cite{disc, rem3}, the crucial point here is the fixed infinite spatial volume $V=\int d^3 x$ occupied by the YMH field configuration. Actually, as it follows from Eq.
 (\ref{masa}), the effective Higgs mass  $m/\sqrt{\lambda}$ is directly proportional to the coupling constant $g$ and thus, as $g$, obeys the    Callan $-$ Symanzik equation \cite{CS} (although with the caution that $m/\sqrt{\lambda}$ is finite when $g\to 0$).

Now we are able to write down explicitly this equation for $m/\sqrt{\lambda}$. Using Eq. (\ref{masa}) and the  Callan $-$ Symanzik equation for the YM coupling constant $g$ (see Eq. (3.62) in the monograph \cite{Cheng})
\be \label{Sym}
\frac{\partial g(t)}{\partial t}=\beta(t), \quad t=\ln \sigma,
\ee
with $\beta(t)$ being the Callan $-$ Symanzik beta-function (the classical look of which is probably known to our readers, so we will not cite it here) and $\sigma$ is the scale in the momentum four-space, we get for  the effective Higgs mass  $m/\sqrt{\lambda}$ the following Callan $-$ Symanzik equation
\be \label{Sym1}
\frac{\partial ( m/\sqrt{\lambda )}}{\partial t}=\frac{<B^2> V}{4\pi}\beta(t).
\ee
At deriving this equation the gauge invariance of $<B^2>$ \cite{Cheng} was utilized.

From this equation we see that if $<B^2>=<B_1^2>=0$ in the $r\to 0$ spatial region where the vortices are located inside the vacuum manifold $R_{{\rm YM}}$ (this is according to (\ref{B1v})) and with the fixed volume $V\to\infty$, the r h s of it {\it is different from zero}. And this, possibly, can be considered as an indicator of the first order phase transition taking place in the YMH BPS monopole vacuum model quantized by Dirac.

\medskip Thus one can consider a diapason $[m_1,\infty]$   in which the effective Higgs mass $m/\sqrt\lambda$ varies  (where $m_1$ of the $O(1/\epsilon(0))$ order is controlled by the Callan $-$ Symanzik equation (\ref{Sym1}) at \linebreak $<B^2>=<B_1^2>=0$)   \footnote{Indeed, infrared QCD effects refer to the interval of distances $[r_h,\infty[$, but any gluonic string confining a quark-antiquark pair near each other cannot stretch to infinite distances; it will tear to a few strings with typical lengths  $\sim$ 1 fm \cite{Cheng}.}. 
Herewith the point $m/\sqrt\lambda=m_1$ {\it is not}  an ultraviolet fixed point for the effective Higgs mass, although $g(0)=0$ and $\partial g / \partial t =0$ simultaneously. It is, of course, a challenge which requires a solution. On the other hand, there is, obviously, a continuous 
(and analytical) renormalization group transformation connecting $m_1$ and the infinite value of mass \cite{LP2} in the infrared confinement region. This allows to interpret the effective Higgs mass $m/\sqrt{\lambda}$ as a {\it Wegner variable} \cite{Kadanoff,Wegner} (this circumstance was noted already in the paper \cite{fund}). 

Another consequence of the above typical mass scale $m_1$ is the possibility to estimate the time $\tau$ at which one can neglect interactions between different topological sectors inside the vacuum manifold $R_{\rm YM}$ in the spatial region (\ref{reg}). Such condition is roughly
\be \label{est}
\tau\gg \hbar/m_1 c^2.
\ee
This is one of the conditions at which the coherent spaces picture is correct for the vacuum manifold $R_{\rm YM}$.

The alternative point of view, i.e. that the above mentioned relaxation time $\tau$ is small, is also permissible. In that case, instead of the pure state describtion and the coherent spaces picture for 
$R_{\rm YM}$, it should go over to the {\it density matrix}  describtion for it.  The same conclusion is correct also in the case when the colliding and annihilation processes for magnetic charges of the same topology (say, $n\neq 0$) \cite{disc, Al.S.} occur very quickly. The said, apparently,  can modify, in some way,  the here discussed theory (for instance, the conclusion  that the physical parameters describing the vacuum manifold $R_{\rm YM}$ are destributed Gaussian). But if we want to build the YMH model involving the first order phase transition (as that represented here and in the references cited), the conditions $<B^2>\neq 0$, $<B_1^2>=0$ still need, presumably, to be carried out. The said requires a lot of further job in order to reconcile the things.

The additional argument in favour of the correctnes of the coherent spaces picture concerning  interaction between (neighboring) topological domains inside the vacuum manifold $R_{\rm YM}$ is disappearing "step voltage" $\phi_2 (n_1+1)-\phi_1 (n_1)$ (with $\phi$'s being, actually, YM potentials belonging to the appropriate topological domains) in the spatial region $r\to 0$, i.e. in the region of the assymptotical freedom of gluons and quarks \cite {disc}  (herewith, as it was pointed out in   \cite {disc}, $<\vec B_1^{n+1}- \vec B_1^{n}>=0$; it is the consequence of Eq. (\ref{B1v})).
But, as it was argued in \cite{disc},  the "step voltage" $\phi_2 (n_1+1)-\phi_1 (n_1)$ is the value of the order  $O(r^\alpha)$ (with $\alpha >0$) at $r\to 0$. Thus the "step voltage" gives a vanishing contribution in the vacuum Hamiltonian for the Minkowskian YMH theory with BPS monopoles quantized by Dirac.

\bigskip Indeed, it is necessary to demand the finite time intervals between "in" and "out" vacuum states belonging to different topologies.    Otherwise, as it was argued in \cite{rem3}, the angular velocity $\dot N(t)$ of collective solid rotations inside the Minkowskian YMH BPS monopole vacuum \cite{rem3,Pervush3},
\be \label{Poi} {\dot N}(t)={\rm const} = (n_{\rm out}-n_{\rm in})/T\equiv \nu/T;  \quad T\equiv   t_{\rm out}- t_{\rm in};    \ee
becomes zero in the $T\to \infty$ limit. This should means that nontrivial vacuum (topological) dynamics disappears in this limit.

The exit from this situation is to consider the finite time interval $\vert T\vert <\infty$ in Eq (\ref{Poi}). This, on the other hand, gives rise \cite {rem3,Logunov} to the problem with the {\it Haag theorem}. According to this theorem, only at $T\to \infty$, it is possible to construct the correct Fock  representation for canonical commutation relations (CCR) between quantum fields (generalized coordinates) $\phi(t,{\bf x}$) and their time derivatives 
(generalized momenta) $\dot \phi (t,{\bf x})$. Instead, one deals with strange quantum states involving "actual" infinite
tower of "bare" particles as $T\neq \infty$.  

The only way to resolve a contradiction now arising (and thus also the possible way solving our problem with disappearing of rotary effects at $T\to \pm \infty $) is to circumvent the Haag theorem.
It turns out that there are odds to do this, but with numerous warnings.

For instance, the Haag theorem does not exclude the existence of the interaction
(Dirac) picture at violating the translational invariance of QFT by introducing the spatial
cut-off \cite{Logunov}.  When the interaction picture exists for the Hamiltonian cutting off in such
a way, we are dealing with the so-called local Fock representation of CCR. However, a
difficult mathematical problem arises in this case with the cut-off removal.

Thus at the current stage of investigations, we can only outline the way out of the $\pm T<\infty$ setting for the Minkowskian YMH physical vacuum model with BPS monopoles and Dirac quantization scheme (in order to include nontrivial topological dynamics there) and an obstacle to this in the form of the Haag theorem on the other hand.
 \par
Indeed, the "coherent state" description (\ref{cohr}) for the vacuum manifold $R_{\rm YM}$ can be applicable even if $T\neq \infty$. 
 For instance, one can neglect the domain walls contribution into the vacuum Hamiltonian. And probably, the only thing which can involve the density matrix description for the vacuum manifold $R_{\rm YM}$ is the effect \cite{disc,Al.S.} colliding and annyhilation of YM (vacuum) modes with topological charges  of an identical topology $n$ (due to "inverting" the sign of the topological number $n$ as such two YM modes interact with a thread Higgs mode with the same topological number $n$), if the time $\tau$, (\ref{est}), is small,
\be \label{est1}
\tau \ll \hbar/m_1c^2.
\ee
To end the discussion about the density matrix description for the vacuum manifold $R_{\rm YM}$, we cite now the explicit look of such density matrix (see e.g. \cite{Levich2}) 
\be \label{dm}
\hat \rho= \frac 1 Z e^{\frac{-H_{\rm vac}}{KT}},
\ee
with $Z$ being the statistic sum:
$$ Z=\sum_n \hat \rho_{nn} \equiv \sum_n w_n = \sum_n e^{-\frac{\epsilon_n}{KT}}\equiv {\rm Tr}e^{-\beta H_{\rm vac}}.$$
Here $\epsilon_n$ are the eigenvalues of the vacuum Hamiltonian $H_{\rm vac}$ and $w_n$ is the probability of the $n^{\rm th}$  eigenvalue of this vacuum Hamiltonian $H_{\rm vac}$ to appear.

Indeed, as it can be seen at examining (\ref{dm}), the density matrix $\hat \rho$ is exponentially suppressed at the temperature $T\to 0$. It is an important conclusion that almost give the answer to our question either the density matrix description is fit for the  vacuum manifold $R_{\rm YM}$ or we should utilize for it the superselection rules picture (\ref{cohr}). Actually, {\it the superselection rules picture is correct at low/zero temperatures of environment}.

\bigskip All the said gives a hope, in spite the first-order phase transition occurring in the Minkowskian YMH BPS monopole model \cite{Pervush1,David2,David3, Pervush2,LP1,LP2} quantized by Dirac, that weak,  $m/\sqrt{\lambda}\to m_1$, and strong, \linebreak $m/\sqrt{\lambda}\to \infty$, coupling regions  
can be connecteed by an  analytical line (referred to as the {\it critical line} in the \linebreak  paper \cite{Kadanoff}) 
\footnote{
As it was analyzed in \cite{disc}, annihilating processes for magnetic charges ${\bf m}\neq 0$ (i.e. appropriate YM BPS monopole modes and excitations over the BPS monopole vacuum) colliding with 
(topologically nontrivial) threads $A_\theta$ can lead (in
a definite time space) to the situation when all such magnetic charges annihilate while Higgs vacuum modes possess arbitrary electric charges (according to the Dirac quantization \cite{Dirac} of the both types of charges). 
In the terminology \cite{Hooft}, one can refer to this as to the Higgs phase (with additional screening ``Higgs'' electric charges by BPS ansatzes \cite{LP1,LP2,Al.S.,BPS}, playing the role of electric formfactors \cite{disc}).  

As it is well known \cite{Hooft}, the Higgs phase is treated as that dual to the confinement phase,  when Higgs vacuum modes are ''magnetic objects'' while quark and gluons are ``electric objects''.

\medskip For the ``ordinary'' Higgs non-Abelian gauge theory the Fradkin-Shenker (Osterwalder-Seiler) theorem takes place \cite{FS}.  It turns out that there are no  transition separating the Higgs and confinement  phases in such theory. But the proof of the  Fradkin-Shenker (Osterwalder-Seiler) theorem losses its validity in the BPS limit \cite{LP1,LP2,Al.S.,BPS} $\lambda\to 0$,  when the Higgs potential decouples from the complete QCD action functional.  

Additionaly, the Fradkin-Shenker (Osterwalder-Seiler) theorem \cite{FS} is valid only in the non-Abelian gauge theory where the Higgs vacuum expectation value $<\Phi>^2$ serves as an order parameter. 

\medskip This creates definite difficulties since  the Higgs and confinement  phases can be now separated each from other. In particular, it can be correctly for the  Minkowskian YMH BPS monopole model \cite{Pervush1,David2,David3, Pervush2,LP1,LP2} quantized by Dirac. Then such ``separation'' will be in an agreement with the first-order phase transition occurring therein but in a definite contradiction with the treatment of the ``effective'' Higgs mass $m/\sqrt{\lambda}$ as a { Wegner variable}. Also $<\Phi>^2$ ceases to be the order parameter in the mentioned model; instead, the value $<B>^2$ for the vacuum ``magnetic'' field $\bf B$ acquires the sense of such a parameter.

The way out from this uncertain situation is, on the author particular opinion, is  in reexamining the  Fradkin-Shenker (Osterwalder-Seiler) theorem in the BPS limit}.

\bigskip

In the recent paper \cite{disc} and in the present study the ways  solving the mass gap problem in the Minkowskian YMH BPS monopole model \cite{Pervush1,David2,David3, Pervush2,LP1,LP2} quantized by Dirac and involving the discrete vacuum geometry (\ref{RYM}) (calling to justify the Dirac fundamental quantization scheme \cite{Dir} applied to this model) are outlined.  Of course, lot of difficulties still remain in this aspect need further study.  For examle, the relation between the   first-order phase transition taking in the Minkowskian YMH model \cite{Pervush1,David2,David3, Pervush2,LP1,LP2} quantized by Dirac and the existence therein the critical line \cite{Kadanoff} connecting weak and strong coupling regions: more exactly, wheter these both  things are compatible each with other or \linebreak not.

 \end{document}